\documentclass[twocolumn,showpacs,preprintnumbers,amsmath,amssymb,showkeys] 
{revtex4}
\usepackage{graphicx}
\usepackage{dcolumn}
\usepackage{bm}
\begin{document}
\title{Dispersion Relations for Bernstein Waves in a Relativistic Pair  
Plasma}
\author{Ramandeep Gill}
  \email{rsgill@phas.ubc.ca}
\author{Jeremy S. Heyl}
  \email{heyl@phas.ubc.ca}
\affiliation{
Department of Physics and Astronomy, University of British Columbia\\
6224 Agricultural Road, Vancouver, BC V6T 1Z1, Canada
}
\date{\today}
\begin{abstract}
   A fully relativistic treatment of Bernstein waves in an
   electron-positron pair plasma has remained too formidable a task
   owing to the very complex nature of the problem. In this article, we
   perform contour integration of the dielectric response function and
   numerically compute the dispersion curves for a uniform, magnetized,
   relativistic electron-positron pair plasma. The behavior of the
   dispersion solution for several cases with different plasma
   temperatures is highlighted. In particular, we find two wave modes
   that exist only for large wavelengths and frequencies similar to the
   cyclotron frequency in a moderately relativistic pair plasma. The
   results presented here have important implications for the study of
   those objects where a hot magnetized electron-positron plasma plays a 
   fundamental role in generating the observed radiation.
\end{abstract}
\maketitle
\section{\label{sec:introduction}Introduction}
Relativistic electron-positron pair plasmas are found in many
astrophysical objects such as neutron star magnetospheres
\cite{RudermanSutherland1975}, relativistic jets and accretion disks
associated with black holes in the centers of active galactic nuclei
\cite{Begelmanetal1984,TakaharaKusunose1985}.  To study the properties
of the pair plasma in these objects it is imperative to employ a fully
relativistic approach.  In the study of magnetized pair plasmas one
finds that a number of oscillation modes exist (see \cite{Melrose1997,
   Gedalinetal2001}). The focus of this study is to investigate the
behavior of Bernstein modes in a uniform, magnetized, relativistic
$e^+e^-$ pair plasma.  Bernstein waves are electrostatic undulations
that are always localized near the electron cyclotron harmonics in an
electron-ion plasma \cite{Bernstein1958}. These waves can propagate
undamped only very close to the plane perpendicular to the static
magnetic field.  Such waves are of great interest since, in an
electron-ion plasma, they strongly interact with the electrons and are
excellent candidates for plasma heating and driving currents as
compared with the electromagnetic (O)rdinary and the e(X)traordinary  
modes
\cite{DeckerRam2006}.

The non-relativistic treatment of Bernstein waves in an electron-ion
plasma is well understood \cite{Swanson2003,KrallTrivelpiece1973,
   BaumjohannTreumann1996,Dougherty1975}. It is the fully relativistic
case that is marred by difficulties such that a closed form analytic
solution is hard to formulate or maybe even impossible. Many workers
have expounded on the fully relativistic treatment of electron
Bernstein waves \cite{Georgiou1996}, the ultrarelativistic case
\cite{Buti1963}, and have successfully obtained approximate dispersion
relations \cite{Saveliev2005,Saveliev2007,Volpe2007}. The Bernstein mode in a
weakly relativistic pair plasma has been investigated by the authors
of \cite{Kestonetal2003} where they have found closed curve dispersion
relations that are remarkably distinct from the classical case. All of
these studies either discuss the fully relativistic case to the point
where no closed form analytic solution is found, or simplify the
analysis by either treating the limiting case only or employ various
approximations that may not yield an entirely correct result.

In this paper, we present a fully relativistic treatment of Bernstein
waves in a pair plasma and provide dispersion curves that highlight
the transition from the weakly to strongly relativistic regime.  The
rest of the article is organized as the following. We derive the key
equations for the dielectric response tensor and describe our
numerical approach to the problem in the following section
(\S~\ref{sec:relat-disp-relat}). In Section~\ref{sec:prop-disp-curv},
we present the main results of this study along with a discussion on
how the solution behaves as a function of the plasma temperature, and
plasma frequency. In the final section (\S~\ref{sec:discussion}) we
highlight some of the important points of the study.

\section{\label{sec:relat-disp-relat}Relativistic Dispersion Relation}

The evolution of the distribution function,
$f(\mathbf{r},\mathbf{p},t)$, of plasma particles in phase space is
governed by the Vlasov equation, which in the momentum representation
is given as
\begin{equation}
\frac{\partial f_s}{\partial t} + \mathbf{v}\cdot\nabla_{\mathbf{r}}  
f_s + q_s
(\mathbf{E} + \mathbf{v} \times \mathbf{B})\cdot\nabla_{\mathbf{p}}  
f_s = 0
\end{equation}
where $s$ indicates different species constituting the plasma. To
investigate the behavior of small amplitude waves with oscillation
periods much smaller than particle collision times, we make the
following assumptions
\begin{eqnarray}
f_s(\mathbf{r},\mathbf{p},t) &=& f_{0s}(p) + f_{1s}(\mathbf{r}, 
\mathbf{p},t) \\
\mathbf{B} &=& \mathbf{B_0} + \mathbf{B}_1 e^{i(\mathbf{k}\cdot 
\mathbf{r}-\omega t)} \\
\mathbf{E} &=& \mathbf{E}_1 e^{i(\mathbf{k}\cdot\mathbf{r}-\omega t)}
\end{eqnarray}
where the subscripts $0$ and $1$ indicate equilibrium and perturbed
functions, respectively.  Furthermore, we restrict the equilibrium
distribution function to only depend on the momentum of the particles
to account for the anisotropy introduced by the ambient static
magnetic field. As a result, we write the Vlasov equation in its
linearized form
\begin{equation}
\frac{d}{dt}f_{1s} = -q_s(\mathbf{E}_1 + \mathbf{v} \times  
\mathbf{B}_1) e^{i (\mathbf{k}
\cdot \mathbf{r} - \omega t)} \cdot \nabla_{\mathbf{p}} f_{0s}
\end{equation}
The main idea here is to calculate the perturbation of the
distribution function by integrating along the unperturbed orbits from
some time in the past, say $t_0$, to the present time $t$.  Next, we
use Ohm's law to write the current density induced by the perturbed
distribution
\begin{equation}
\mathbf{J} = \sum_s \frac{q_s}{m_s} \int \mathbf{p}_s f_{1s} 
(\mathbf{r},\mathbf{p},t) d^3p = \sum_s  
\underline{\underline{\sigma}}_s \cdot \mathbf{E_1}
\end{equation}
This enables us to write the effective dielectric permittivity tensor
in terms of the conductivity tensor $\underline{\underline{\sigma}}$
\cite{Swanson2003}
\begin{equation}
\underline{\underline{\epsilon}}(\omega,\mathbf{k}) =  
\epsilon_0\left(I-\frac{\underline{\underline{\sigma}}}{i\omega_s  
\epsilon_0}\right)
\end{equation}
In the fully relativistic approximation, the energy and linear
momentum of particles in the rest frame of the plasma are given as
\begin{eqnarray}
&& \mathcal{E} = \gamma mc^2 = \sqrt{p^2 c^2 + m^2 c^4} \\
&& \mathbf{p} = \gamma m \mathbf{v} \nonumber
\end{eqnarray}
where the Lorentz factor is given in terms of the momentum as
\begin{equation}
\gamma = \left( 1 + \frac{p^2}{m^2 c^2} \right)^{\frac{1}{2}}
\end{equation}
The complete details of the rest of the calculation can be found in
various monographs and textbooks on plasma waves (see
\cite{Swanson2003,BaumjohannTreumann1996}), and we only provide the
salient points of the derivation in what follows.  We adopt
$\mathbf{B}_0=B_0\hat{z}$ for the equilibrium magnetic field and
restrict both the wave vector and the perturbed electric field to be
$\mathbf{k}=k_{\perp}\hat{x}$ and $\mathbf{E}_1=E_1\hat{x}$ as
dictated by the purely electrostatic mode where the perturbed magnetic
field $\mathbf{B}_1$vanishes.  After some mathematical manipulations
we arrive at the relativistic dielectric tensor
\begin{equation}
\underline{\underline{\epsilon}}(\omega,\mathbf{k}) = 
\left(\begin{array}{ccc}
\epsilon_{xx} & \epsilon_{xy} & 0 \\
-\epsilon_{xy} & \epsilon_{yy} & 0 \\
0 & 0 & \epsilon_{zz}
\end{array}\right)
\end{equation}
where the different components have been summed over both species, 
$e^+$ and $e^-$, of the pair plasma
\begin{eqnarray}
\epsilon_{xx} &=& \epsilon_0 \left[ 1 + \frac{4 \pi q^2
    m^2}{k_{\bot}^2 m \epsilon_0}  \int \left\{ \frac{\pi \gamma a}{\sin
    \pi \gamma a} J_{\gamma a} (\xi) J_{- \gamma a} (\xi) \right.  
\right. \nonumber \\
    & & \left. \left. - 1 \right\} \frac{\gamma}{p_{\bot}}  
\frac{\partial f_0 (p)}{\partial p_{\bot}} p^2
    \sin \theta dpd \theta \right] \nonumber \\
\epsilon_{yy} &=& \epsilon_0  \left[ 1 + \frac{4 \pi
    q^2}{\omega m \epsilon_0}  \int  \frac{p_{\bot}}{\omega_c}
    \frac{\partial f_0 (p)}{\partial p_{\bot}}  \left\{ \frac{\pi} 
{\sin \pi
    \gamma a} \right. \right. \nonumber \\
    & & \left. \left. \times J'_{\gamma a_{}} (\xi) J'_{- \gamma a}  
(\xi) + \frac{a}{\gamma
    \xi^2} \right\} p^2 \sin \theta dpd \theta \right] \nonumber
\end{eqnarray}
\begin{eqnarray}
\epsilon_{zz} &=& \epsilon_0 \left[ 1 + \frac{4 \pi q^2}{\omega m
    \epsilon_0 \omega_c}  \int p_{\parallel}  \frac{\partial f_0
    (p)}{\partial p_{\parallel}}  \right. \nonumber \\
    & & \left. \times \frac{\pi J_{\gamma a} (\xi) J_{- \gamma a}  
(\xi)}{\sin \pi \gamma a}
    p^2 \sin \theta dpd \theta \right] \nonumber \\
\epsilon_{xy} &=& i \frac{4 \pi q^2 m^2}{m
    k_{\bot}^2}  \int \frac{\gamma}{p_{\bot}}   
\frac{\partial f_0
    (p)}{\partial p_{\bot}} p^2 \sin \theta dpd \theta
\end{eqnarray}
In the above, $\parallel$ and $\perp$ subscripts denote components
parallel and perpendicular to the equilibrium magnetic field,
$J_{\gamma a}(\xi)$ is the Bessel function of non-integer order,
$J'_{\gamma a}(\xi)$ is the derivative of the Bessel function with
respect to $\xi$, where $\xi=\frac{k_{\perp} p_{\perp}}{q B}$, and
$a=\frac{\omega}{\omega_c}$ with $\omega_c$ denoting the
non-relativistic cyclotron frequency.  Finally, one finds the
dispersion relation, $\omega = \omega(\mathbf{k})$, by setting the
dielectric response function to zero,
\begin{equation}
\epsilon(\omega, \mathbf{k}) = \mathbf{k} \cdot  
\underline{\underline{\mathbf{\epsilon}}}(\omega, \mathbf{k}) \cdot  
\mathbf{k} = 0
\end{equation}
which in our case simply picks out the $\epsilon_{xx}$ component.  To
keep the treatment fully relativistic we adopt the
Maxwell-Boltzmann-J\"utner distribution function
\cite{Swanson2003,Schlickeiser1998},
\begin{equation}
f_0 (p) = (4 \pi m^3 c^3)^{- 1}  \frac{\eta}{K_2 (\eta)} e^{- \eta  
\gamma}
\end{equation}
where
\begin{equation}
\eta \equiv \frac{mc^2}{k_B T}
\end{equation}
is the ratio of the rest mass energy of the particles to that of their  
thermal
energy, and $K_2$ is the modified Bessel function of the second kind  
and of order two.
Also, the equilibrium plasma distribution has been normalized to unity
\begin{equation}
1 = n_0 = \int f_0(p) d^3p
\end{equation}
Taking the derivative of $f_0(p)$ with respect to $p_{\perp}$ and  
$p_{\parallel}$ yields
\begin{equation}
\frac{\partial f_0}{\partial p_{\spadesuit}} = -\frac{\eta^2}{4\pi m^5  
c^5 K_2(\eta)}\frac{p_{\spadesuit}}{\gamma}e^{-\eta\gamma}
\end{equation}
where $\spadesuit$ can be replaced by either $\parallel$ or $\perp$  
components.
At this point, we can carry out the integration over the polar angle  
and by defining
$\beta=\frac{k_{\perp} p}{q B}$ we can write $\epsilon_{xx}$ as the  
following
\begin{eqnarray}
&& \epsilon_{xx} = \epsilon_0 \left[ 1 - \frac{\omega_p^2 \eta^2} 
{k_{\bot}^2 m^3 c^5 K_2(\eta)} \int_0^{\infty} p^2 e^{- \eta \gamma}  
\right. \\
&& \left. \times \int_0^{\pi} \left\{ \frac{\pi
   \gamma a}{ \sin \pi \gamma a} J_{\gamma a} (\beta \sin \theta) J_{- 
\gamma a} (\beta \sin \theta) - 1 \right\} \sin \theta d \theta dp  
\right] \nonumber
\end{eqnarray}
where the non-relativistic plasma frequency is defined as
\begin{equation}
\omega_p^2=\frac{n_0e^2}{m\epsilon_0}
\end{equation}
Next we use the following Bessel function identity \cite{Swanson2003}
\begin{eqnarray}
&& \int_0^{\pi} \sin \theta J_a (b \sin \theta) J_{- a} (b \sin  
\theta) d \theta \\
&&  = \frac{2 \sin \pi a}{\pi a} {_2F_3}  \left( \frac{1}{2}, 1 ;  
\frac{3}{2}, 1 - a, 1 + a ; - b^2 \right) \nonumber
\label{eq:besselid}
\end{eqnarray}
to express the integral over the polar angle in terms of a  
hypergeometric function, and redefine all constants and variables to  
make $\epsilon_{xx}$ dimensionless
\begin{eqnarray}
&& \hat{p} = \frac{p}{mc} \ \ \  \hat{k}_{\bot} = \frac{k_{\bot} c} 
{\omega_0} \ \ \
   \hat{\omega} = \frac{\omega}{\omega_0} \ \ \ \\
&&  \hat{\omega}_p =
   \frac{\omega_p}{\omega_0} \ \ \  \beta = \hat{\beta} =  
\hat{k}_{\bot} \hat{p} \ \ \ a =
   \hat{\omega} \nonumber
\end{eqnarray}
Then, we can write $\epsilon_{xx}$ as the following where the integral  
now is
just over the dimensionless momentum $\hat{p}$,
\begin{eqnarray}
\label{eq:exx}
&& \epsilon_{xx} = \epsilon_0 \left[ 1 - \frac{2
    \hat{\omega}_p^2 \eta}{\hat{k}_{\bot}^2} \left\{ \frac{\eta}{K_2  
(\eta)}
    \int_0^{\infty} \hat{p}^2 e^{- \eta \gamma} \right. \right. \\
&& \left. \left. \times {_2F_3}
    \left( \frac{1}{2}, 1 ; \frac{3}{2}, 1 - \gamma \hat{\omega}, 1 +  
\gamma
    \hat{\omega} ; - \hat{\beta}^2 \right) d\hat{p} - 1 \right\}  
\right] \nonumber
\end{eqnarray}
In writing this equation we have made use of the following integral  
identity
\begin{equation}
\int_0^{\infty} \hat{p}^2 e^{-\eta \gamma} d\hat{p} = \frac{K_2(\eta)} 
{\eta}
\end{equation}
The integrand in Eq. (\ref{eq:exx}) consists of a hypergeometric function
which is singular for $1-\gamma \hat{\omega} = -n$ for integer $n = 0,  
1, 2, \ldots ,$.
This singular behavior, as we shall see, is associated to the  
phenomenon of cyclotron resonance
where plasma particles are in resonance with the wave at the cyclotron  
harmonics. Also, this very resonance poses
a real challenge for any numerical computation of the integral and has  
to be dealt with using
advanced numerical techniques. Since our interest lies in finding the
oscillation frequency $\hat{\omega}$ as a function of the wavenumber $ 
\hat{k}_{\perp}$,
it is clear from Eq. (\ref{eq:exx}) that this operation is explicitly  
nonlinear. Therefore, one is left with an
exercise of root finding for a given $\hat{\omega}$.
Alternatively, one could simplify the analysis
by making some approximation. However, by adopting such methodology  
one risks
losing the subtleties of the solution and may obtain something that is  
not
entirely correct, as we show in the weakly relativistic case. We  
remain optimistic
and decide to compute the dispersion relation using a brute force  
method, that is
by simply integrating Eq. (\ref{eq:exx}).
\subsection{\label{sec:numerical-approach}Numerical Approach}
\begin{figure}[h]
\includegraphics{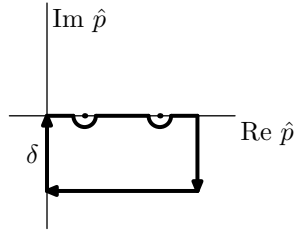}
\caption{Contour of integration in the complex $\hat{p}$-plane used to  
avoid
singular points of the integrand. Only the general case is shown here  
to guide
the reader.}
\label{fig:contour}
\end{figure}

As the Lorentz factor is a function of momentum, the integrand remains  
singular
over the domain of integration. A workaround for avoiding the singular  
points on
the real axis is by analytically continuing the momentum to the  
complex domain.
We can easily shift the integration contour below the real axis by  
writing
$p \rightarrow p - i \delta$
where $\delta$ is reasonably small. Ideally, one would like to keep  
the contour on
the real axis but go below the singular points to avoid divergence  
while following
the Landau prescription \cite{Landau1946}. A similar result can be  
achieved by closing the contour of
integration in the lower half of the complex plane as shown in 
Fig. \ref{fig:contour}. 
Here we are interested in finding the 
principal value of the integral in Eq. (\ref{eq:exx}) which is 
given by the following,
\begin{equation}
\mathrm{p.v.} \int_0^\infty d\hat{p}\ f(\hat{p}) = i\pi\sum_{\mathrm{Residue}} R_0 - (I_2+I_3+I_4)
\label{eq:pv}
\end{equation}
where
\begin{eqnarray}
&& I_2 = \int_\infty^{\infty - i\delta} d\hat{p}\ f(\hat{p}) \approx 0 \nonumber \\
&& I_3 = \int_{\infty - i\delta}^{-i\delta} d\hat{p}\ f(\hat{p}) \\
&& I_4 = \int_{-i\delta}^0 d\hat{p}\ f(\hat{p}) \nonumber 
\label{eq:contoureq}
\end{eqnarray}
and $R_0$ is the residue from the poles on the Re($\hat{p}$) axis. 
In the above, for $I_2$ we note that the integrand 
vanishes sufficiently rapidly for large $\hat{p}$ due to the  
exponential. When $\hat{\omega}$ and $\hat{k}_\perp$ are kept real, one 
finds that $R_0$ is also real which makes the first term on the right in Eq. (\ref{eq:pv}) 
purely imaginary. However, the second term has both real and imaginary components, 
and the negative sign is indicative of the clockwise sense of the contour. Since the 
left hand side is real then so must be the right, which suggests that the imaginary components 
cancel each other and yields the following result
\begin{equation}
\mathrm{p.v.} \int_0^\infty d\hat{p}\ f(\hat{p}) = - \mbox{Re}(I_3+I_4)
\end{equation}

Although it appears that with the given prescription one can have
an arbitrarily large $\delta$, we find that the integral does start to  
lose its
accuracy as $\delta$ approaches unity. Therefore, we set $\delta=0.1$  
for all
numerical computations. To compute the integral numerically over the hypergeometric function
we used Mathematica (V.6) for it is capable of calculating generalized
hypergeometric functions. The poles of the integrand were dealt with
by employing a globally adaptive integration routine available in
Mathematica. To speed up the integration over the singular points we
used the Double Exponential Quadrature singularity handler built into
Mathematica's integration routine.

\section{\label{sec:prop-disp-curv}Properties of Dispersion Curves}

\begin{figure}[h]
\includegraphics[scale=0.7]{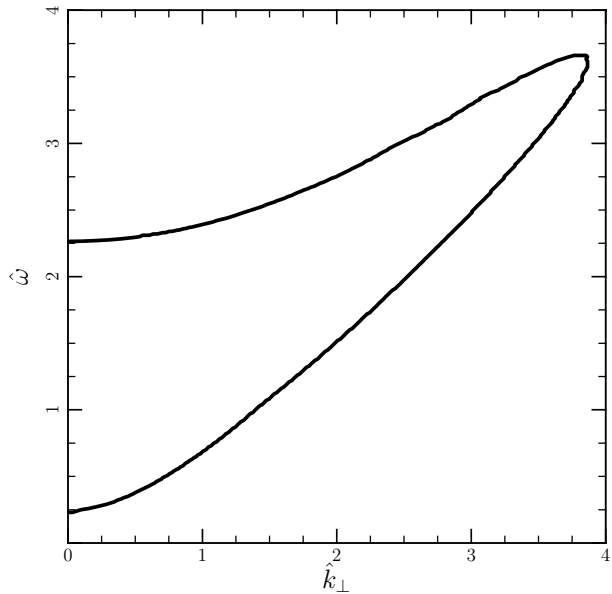}
\caption{Dispersion curves for a relativistic plasma with $\eta=1$, $ 
\hat{\omega}_p=3$.
Hatted variables are expressed in terms of the non-relativistic  
cyclotron frequency
$\omega_c$. See text for more detail.}
\label{fig:wp3eta1}
\end{figure}

\begin{figure}[h]
\includegraphics[scale=0.7]{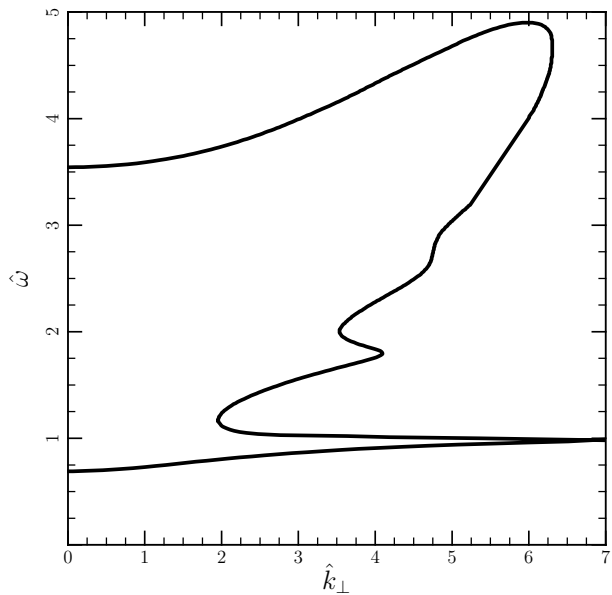}
\caption{Dispersion curves for a relativistic plasma with $\eta=5$, $ 
\hat{\omega}_p=3$.
Hatted variables are expressed in terms of the non-relativistic  
cyclotron frequency
$\omega_c$. See text for more detail.}
\label{fig:wp3eta5}
\end{figure}

\begin{figure}
\includegraphics[scale=0.7]{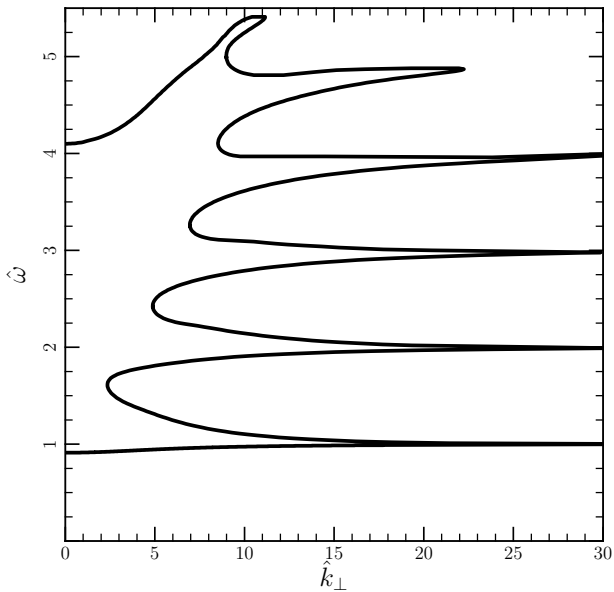}
\caption{Dispersion curves for a relativistic plasma with $\eta=20$, $ 
\hat{\omega}_p=3$.
Hatted variables are expressed in terms of the non-relativistic  
cyclotron frequency
$\omega_c$. See text for more detail.}
\label{fig:wp3eta20}
\end{figure}

In the non-relativistic case of the electron Bernstein modes in an
electron-ion plasma, one finds that there are no wave modes below the
first harmonic. The dispersion curves above the hybrid frequency are
all bell shaped with local maxima corresponding to stationary
modes. Furthermore, band gaps are present above the hybrid frequency
between each dispersion curve. It is not at all surprising to say that
the picture is remarkably different in the relativistic pair plasma
scenario. We plot the dispersion curves for a relativistic pair plasma
in Fig. \ref{fig:wp3eta1}, \ref{fig:wp3eta5}, \ref{fig:wp3eta20}
for different values of $\eta = 1, 5, 20$, respectively. We assume a
plasma frequency of $\hat{\omega}_p=3$ for these plots.

The dispersion relation for a moderately relativistic pair plasma,
shown in Fig. \ref{fig:wp3eta1}, clearly has two wave modes. This is
further accompanied by the existence of two stationary modes with 
vanishing group velocity ($d \hat{\omega} / d \hat{k}_\perp = 0$) 
at two distinct oscillation
frequencies for $\hat{k}_{\perp} = 0$. Also, above the higher
stationary mode there are two wavenumber solutions for a given
$\hat{\omega}$. This behavior persists in the case of a mildly
relativistic pair plasma, shown in Fig. \ref{fig:wp3eta5}. However,
one sees some drastic changes in the shape of the dispersion curves as
the particles lose their energy. We readily notice the appearance of
the curve near the cyclotron fundamental frequency. This marks the
onset of the cyclotron resonance where the plasma particles oscillate
at the same frequency as the perturbing electrostatic wave. Although
not very significant at this point, the higher harmonic resonances
also start to emerge. Furthermore, the dispersion relation now extends
to higher wavenumbers and the turnover from the lower wave mode into
the upper mode is not as sharp as it was in the previous case. We also
notice a shift in the stationary modes, and we discuss this point in a
later section.

The authors of \cite{Kestonetal2003} reported the dispersion relation
for a weakly relativistic pair plasma. They found island shaped curves
occurring mostly between the cyclotron harmonics.  We find a similar
solution for the weakly relativistic case, shown in
Fig. \ref{fig:wp3eta20}, but with a few exceptions. Firstly, the
dispersion curves are no longer closed but extend to higher
wavenumbers as they approach cyclotron harmonics. This behavior is
very similar to what we observe in the case of a non-relativistic
electron-ion plasma. However, in the pair plasma case the solution
does not extend to an infinitely large wavenumber, and there are not
infinitely many wave modes as we increase $\hat{\omega}$.  There
appears to be a cutoff in frequency, the point where the overturn
takes place, beyond which there does not exist any solution. Secondly,
there exists a wave mode below the fundamental cyclotron harmonic
which was absent in the solution provided in \cite{Kestonetal2003}. In
fact, we find that a solution below the cyclotron fundamental exists
for all cases, regardless of $\eta$, for $\hat{\omega}_p=3$ as we show
below. This, again, is in contrast with the non-relativistic
electron-ion plasma where there is no wave mode below the first
harmonic.

In comparison to the moderately relativistic case, the weakly
relativistic case is richer in its behavior as well as much more
structured. The former only has two frequency modes for a given
wavelength, namely a high and a low mode, and the latter has many.
Moreover, for a moderately relativistic pair plasma none of the wave
modes found in between the two stationary points (discussed below)
extend to infinitely small wavelengths. In fact, there is a limiting
wavelength above which these modes exist.  As reported in
\cite{Kestonetal2003}, one can find similar, although much less
severe, imperfections in the graphics produced by the contouring
algorithm. The culprit here is the non linearity of the equation from
which the solutions are obtained.
\begin{figure}[h!]
\includegraphics[scale=0.7]{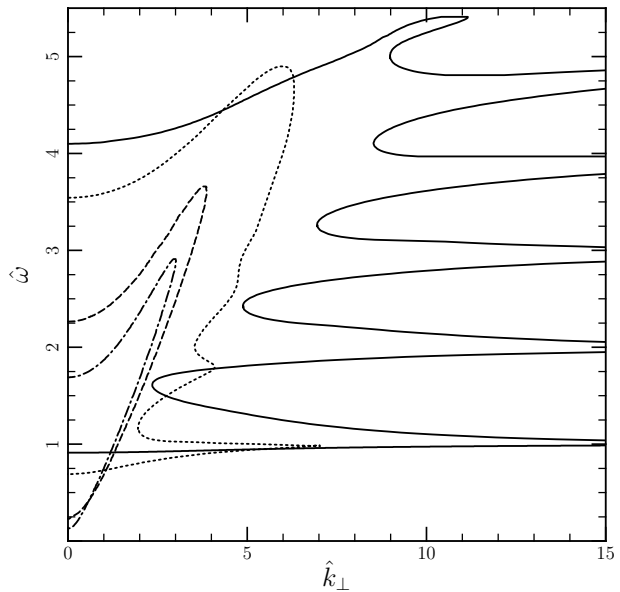}
\caption{Dispersion curves for different values of $\eta$ with $ 
\hat{\omega}_p=3$. (Solid) The weakly relativistic case with $ 
\eta=20$, (Dots) the mildly
relativistic case with $\eta=5$, and the two strongly relativistic  
cases: (a) (Dash) with $\eta=1$ (b) (Dot-dash) with $\eta=0.5$}
\label{fig:wp3all}
\end{figure}
\subsection{\label{sec:relativistic-effects}Relativistic effects}
As the plasma particles become strongly relativistic, with decreasing
$\eta$, the dispersion curves undergo drastic changes. We now plot all
of the dispersion curves shown earlier onto a single plot and analyze
the progression from the weakly relativistic case to the strongly
relativistic one (see Fig. \ref{fig:wp3all}). The Bernstein waves are
strongly absorbed near the cyclotron harmonics in the weakly
relativistic limit where $\eta \gg 1$.  This effect is the strongest
at the first cyclotron harmonic at which point the phase velocity of
the wave $v_{ph}\rightarrow 0$ and the wave loses all of its energy in
heating up the pair plasma. At higher harmonics the same phenomenon is
repeated, however with decreased efficiency. Upon increasing the
thermal speed of the plasma particles, the cyclotron resonances become
much less pronounced and start to disappear completely. In a hot
magnetized pair plasma no resonant interaction between the Bernstein
wave and the plasma occurs, and the solution occupies only a small
region of the $\hat{\omega}-\hat{k}_{\perp}$ space.

\subsection{\label{sec:behav-small-hatk_p}Behavior for Small $ 
\hat{k}_{\perp}$}

Unlike the non-relativistic electron Bernstein waves, the behavior of
the dispersion curves in the limit of vanishing wavenumber is not so
straightforward. The hypergeometric function in the integrand of
Eq. (\ref{eq:exx}) can also be written in the form of an infinite power
series \cite{Swanson2003}
\begin{eqnarray}
&& {_2}F_3(a_1,a_2;b_1,b_2,b_3;x) = \frac{\Gamma (b_1) \Gamma (b_2)  
\Gamma (b_3)}{\Gamma (a_1) \Gamma (a_2)} \\
&& \times \sum_{m=0}^{\infty}\frac{\Gamma (a_1+m) \Gamma (a_2+m)} 
{\Gamma (b_1+m) \Gamma (b_2+m) \Gamma( b_3+m)}
\frac{x^m}{m!} \nonumber
\end{eqnarray}
We see that for $x\rightarrow 0$ only the $m=0$ term has the dominant
contribution. In that case, we find that the hypergeometric function
tends to unity. Consequently, $\epsilon_{xx}$ loses its 
dependence on $\hat{\omega}$ because the only place $\hat{\omega}$ appears 
in Eq. (\ref{eq:exx}) is inside the hypergeometric function. Moreover, 
this suggests that $\hat{\omega}(\hat{k}_\perp)$ is constant in the domain 
where $\hat{k}_\perp \ll 1$ and Eq. (\ref{eq:exx}) fails to
describe the behavior of the dispersion relation for vanishing
$\hat{k}_{\perp}$. In fact, one has to keep terms in the infinite sum
up to order $m=2$ to obtain any non-trivial solution. However, upon
doing so one finds that the integral, again, is extremely non-trivial
and its solution cannot be expressed analytically. This is problematic 
because the dispersion solution is no longer a smooth function and may 
become discontinuous for small $\hat{k}_\perp$.

Nevertheless, this problem can be resolved easily. Upon cursory
inspection of the integrand one finds that it is quadratic in
$\hat{k}_{\perp}$, making it symmetric under the transformation
$\hat{k}_{\perp}\rightarrow -\hat{k}_{\perp}$.  Furthermore, we demand 
that the dispersion relation be continuous at $\hat{k}_\perp = 0$, 
thus, we employ polynomial interpolation to determine the
$y$-intercept. In all three figures we use a polynomial of order 2 or
4, depending on the shape of the curve, to determine the stationary
points for vanishing wavenumber.

\subsection{\label{sec:behav-large-hatk_p}Behavior for Large $ 
\hat{k}_{\perp}$}

In the weakly relativistic case, one finds wave-particle resonances
occurring at cyclotron harmonics. These resonances extend to large
values of $\hat{k}_{\perp}$ but not to infinity. Intuitively, one may
argue, by looking at the rest of the dispersion solution, that such a
behavior is expected as the dispersion curves start and end at
$\hat{k}_{\perp}=0$.  The plot remains connected over the whole
domain, as is evident in the strongly relativistic case, and the
resonances only extend to some maximum value of the wavenumber
$\hat{k}_{\perp}^\mathrm{Max}$. This hypothesis can be ascertained by
computing the integral in Eq. (\ref{eq:exx}) for $\hat{k}_{\perp} \gg 1$,
however for large values of $\hat{k}_{\perp}$ the calculation becomes
very computationally expensive.

Ideally, one would like to find the asymptotic behavior of the
hypergeometric function in Eq. (\ref{eq:exx}) to simplify the problem. The
asymptotic behavior of the hypergeometric function can be gleaned by
asymptotically expanding the Bessel function in the integral
representation of the hypergeometric function given in
Eq. (\ref{eq:besselid}). The Bessel function expansion for large arguments
is given as \cite{GradshteynRyzhik2007}
\begin{equation}
J_{\pm \nu}(z) = \sqrt{\frac{2}{\pi z}} \cos\left(z \mp \frac{\pi} 
{2}\nu - \frac{\pi}{4} \right) + \mathcal{O}\left(z^{-3/2}\right)
\end{equation}
Plugging this back into Eq. (\ref{eq:besselid}) and carrying out the
integral over the polar angle yields,
\begin{eqnarray}
&& \frac{2}{\pi b} \int_0^{\pi} d\theta \cos\left(b\sin\theta -  
\frac{\pi}{2}a - \frac{\pi}{4}\right)
\cos\left(b\sin\theta + \frac{\pi}{2}a - \frac{\pi}{4}\right)  
\nonumber \\
&& = \frac{\cos(a\pi) + H_0(2b)}{b}
\end{eqnarray}
where $H_0$ is the Struve function of order 0. For our specific case,
$b = \hat{k}_{\perp} \hat{p}$ and $a = \gamma \hat{\omega}$, and upon
substitution of this result into the dielectric response function we
find
\begin{eqnarray}
&& \epsilon_{xx} = \epsilon_0\left[1-\frac{2\hat{\omega}_p^2\eta} 
{\hat{k}_{\perp}^2}\left\{\frac{\eta}{2\hat{k}_{\perp}K_2(\eta)}
\int_0^{\infty}\hat{p}e^{-\eta\gamma} \right. \right. \nonumber \\
&& \left. \left. \times \left(\frac{\pi\gamma\hat{\omega}}{\sin \pi 
\gamma\hat{\omega}}\right)[\cos \pi\gamma\hat{\omega}
+ H_0(2\hat{k}_{\perp}\hat{p})]d\hat{p} - 1\right\}\right]
\label{eq:simpleexx}
\end{eqnarray}
This integral again has a similar singularity for $\gamma\hat{\omega}
= n$ for integer $n$. We can further simplify this equation by noting
the leading order behavior of the Struve function in the limit
$b\rightarrow\infty$, which goes like
$H_0(b)\propto\frac{1}{\sqrt{b}}$. Then, in this limit the dominant
term in Eq. (\ref{eq:simpleexx}) is the cosine term. Although this step is
not justifiable given the limits of integration where the integrand is
evaluated for $\hat{p} \ll 1$, this does not modify the overall
behavior of the dielectric response function in the large
$\hat{k}_{\perp}$ limit. Next, we define
\begin{equation}
\mathcal{I}(\eta,\hat{\omega}) = \pi \hat{\omega} \int_0^{\infty} d 
\hat{p}\ \gamma\hat{p}e^{-\eta\gamma} \cot(\pi\gamma\hat{\omega})
\label{eq:I}
\end{equation}
and solve for $\epsilon_{xx}=0$. After some rearrangement of terms, we
arrive at a cubic equation which we then solve for $\hat{k}_{\perp}$
\begin{equation}
\hat{k}_{\perp}^3 + 2\hat{\omega}_p^2\eta\hat{k}_{\perp} -  
\frac{\hat{\omega}_p^2\eta^2}{K_2(\eta)}\mathcal{I}(\eta,\hat{\omega})  
= 0
\end{equation}
To do the integral in Eq. (\ref{eq:I}) we again employ the same contour
integration scheme as was done earlier (see Fig. \ref{fig:contour}).
For $\eta=20$ and $\hat{\omega}=1$, which is the strongest resonance
of all occurring at higher cyclotron harmonics, we find a maximum
wavenumber $\hat{k}_{\perp}^\mathrm{Max} \sim 72$. We find values of
the same order for resonances at higher harmonics as well. This limit
shows the maximum value of $\hat{k}_{\perp}$ for which a solution exists using
the simplified dielectric tensor (appropriate for large values of
$\hat{k}_{\perp})$.  The actual limits to the wavenumber in the realistic
dispersion relation are typically lower.  This exercise demonstrates
that the resonances do not extend to infinitely small wavelengths,
that there is some cutoff at $\hat{k}_{\perp} \sim
\hat{k}_{\perp}^\mathrm{Max}$, and the dispersion curves remain
connected over the whole domain. More importantly, one must not forget
that this estimate is particularly inaccurate for the strongly
relativistic case where the solution exists for only modest values of
$\hat{k}_{\perp}$.

\subsection{\label{sec:stationary-modes}Stationary Modes}

\begin{figure}
\includegraphics[scale=0.7]{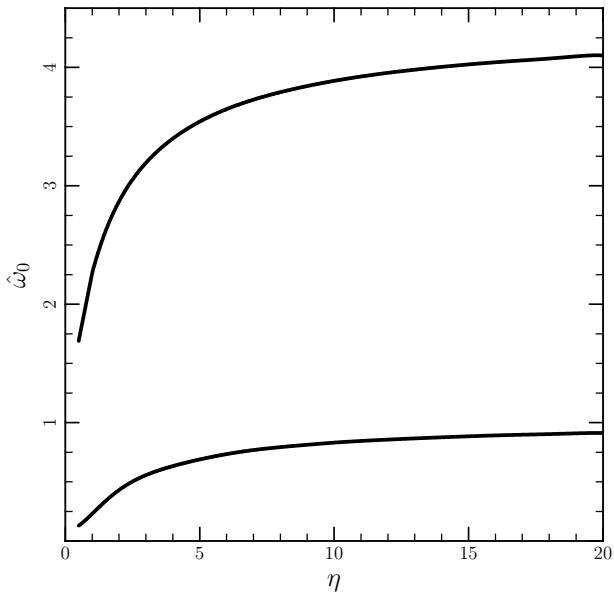}
\caption{Stationary modes at vanishing $\hat{k}_{\perp}$ as a
   function of non-dimensional reciprocal temperature parameter $\eta$.
   The upper curve corresponds to the upper stationary mode and the
   lower curve to the lower stationary mode. For this plot we assume
   $\hat{\omega}_p=3$.}
\label{fig:spVeta}
\end{figure}

There are two stationary modes ($v_g=d\hat{\omega}/d\hat{k}_{\perp}=0$) present for
vanishing $\hat{k}_{\perp}$ in all the cases shown above. We plot the
evolution of both stationary points as a function of $\eta$ in
Fig. \ref{fig:spVeta}. For $\omega_p=3\omega_c$ the upper stationary
mode remains above the cyclotron fundamental and the lower stationary
mode remains below it for all $\eta$. Contrastingly, in the
electron-ion case a solution exists for vanishing $\hat{k}_{\perp}$ at
all cyclotron harmonics, except at the fundamental. It appears that
the upper stationary mode turns into the hybrid resonance given by
\begin{equation}
\hat{\omega}_H^2 = \hat{\omega}_p^2 + 1
\end{equation}
in an electron-ion plasma.

\begin{figure}
\includegraphics[scale=0.7]{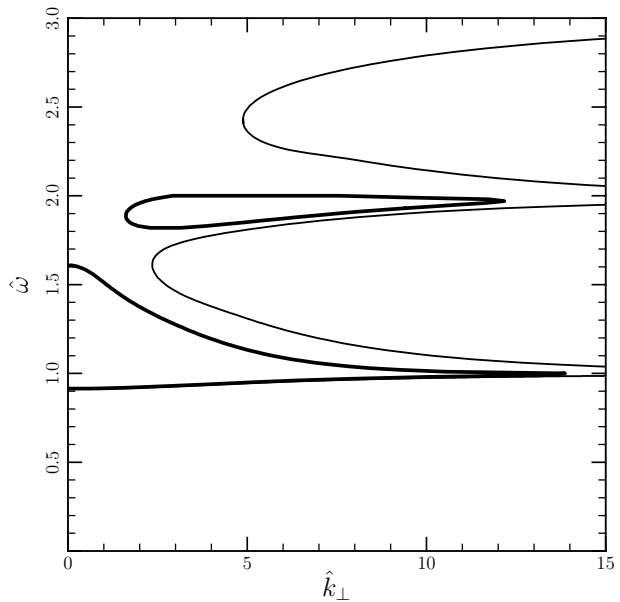}
\caption{Dispersion curves at two different plasma frequencies with $ 
\eta=20$. (a) For $\hat{\omega}_p=3$, and (b) (bold) for $ 
\hat{\omega}_p=1$.}
\label{fig:wp1eta20}
\end{figure}

This picture is slightly modified as we lower the plasma frequency so
that it equals the cyclotron frequency, $\hat{\omega}_p = 1$, while
remaining in the weakly relativistic limit with $\eta=20$ (see
Fig. \ref{fig:wp1eta20}). We still find those two stationary points
and the strong resonance at the cyclotron fundamental, however, the
rest of the plot has disappeared, and, interestingly, been replaced by
a single closed curve. By comparing the present case to the one
treated previously, with $\hat{\omega}_p=3$, we find that upon
decreasing the plasma frequency the lower branch of the dispersion
curve in the first harmonic band separates from the upper
branch. Also the upper branch in the first harmonic band connects to
the lower branch in the second harmonic band. This is the first
incidence where a closed curve solution, like the ones found by the
authors of \cite{Kestonetal2003}, has appeared in our
analysis. Furthermore, the number of stationary modes are now double
of what was observed previously.
\begin{figure}
\includegraphics[scale=0.7]{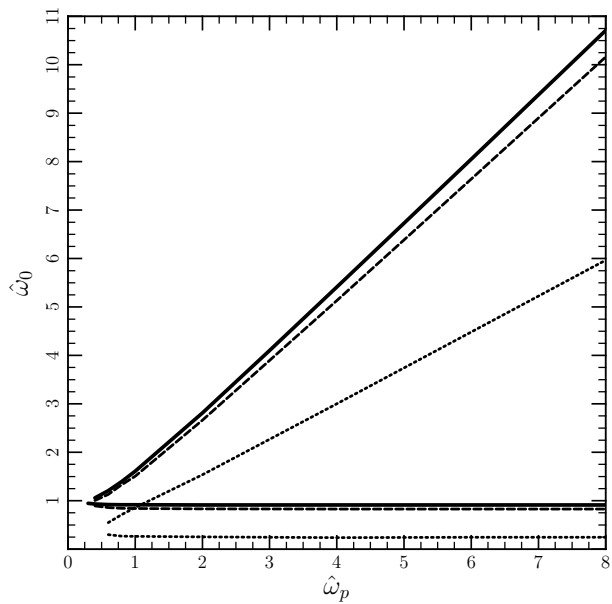}
\caption{Stationary modes as a function of the plasma frequency at  
different values of the reciprocal temperature parameter $\eta$. (a)  
(Solid) For $\eta=20$,
(b) (Dash) for $\eta=10$, and (c) (Dot) for $\eta=1$.}
\label{fig:spVwpeta}
\end{figure}

We also provide a plot of the position of stationary points in
frequency as a function of the plasma frequency for different values
of $\eta = 1, 5, 20$ (see Fig. \ref{fig:spVwpeta}).  As the slope of
the curves varies with a change in $\eta$, the upper hybrid frequency,
if one was to associate the upper stationary mode with it, necessarily
depends on the temperature of the plasma, as shown in
Fig. \ref{fig:spVeta}. Interestingly, in the case of a hot pair plasma
($\eta \leq 1$) we find that the upper stationary mode extends below
the cyclotron fundamental for $\hat{\omega}_p < 1$. Thus, an
underdense ($\omega_p < \omega_c)$ strongly relativistic pair plasma
does not have any Bernstein wave modes above the cyclotron
fundamental. Another consequence of this situation is that the two
wave modes exist for only small wavenumbers and therefore for
extremely large wavelengths. We see that for a given plasma
temperature, the upper stationary mode depends linearly on the plasma
frequency for $\hat{\omega}_p > 1$. On the other hand, the lower mode
remains constant.
\section{\label{sec:discussion}Discussion}
In this article, we investigate the behavior of Bernstein waves in a
uniform, magnetized, relativistic electron-positron pair plasma and
provide dispersion curves for different values of the non-dimensional
reciprocal temperature. The dispersion solutions in all cases are
found to be remarkably different than the Bernstein modes found in the
non-relativistic electron-ion case. For a moderately relativistic pair
plasma we find two Bernstein wave modes accompanied by two stationary
modes for vanishing wavenumber. We do not find closed curve solutions
\cite{Kestonetal2003} in all cases but one where the plasma frequency
equals the cyclotron frequency in the weakly relativistic limit.

As stated earlier, the Bernstein waves in a non-relativistic
electron-ion plasma propagate undamped in the direction orthogonal to
the equilibrium magnetic field. This might not be true for such waves
in a relativistic pair plasma. These waves were found to be very
weakly damped in a weakly relativistic pair plasma
\cite{LaingDiver2005}. In this article, we only report the real
component of $\hat{\omega}$ as the imaginary component is quite 
non-trivial to calculate for the following reason. With 
$\hat{\omega}$ complex Eq. (\ref{eq:contoureq}) does not hold since 
the principal value of the integral is no longer purely real. As a 
result, one must calculate the residue $R_0$ which itself 
is a difficult task as the integrand in Eq. (\ref{eq:exx}) cannot 
be easily transformed into the following form where the singularity 
arises due to a simple pole of order $m$,
\begin{equation}
f(\hat{p}) = \frac{g(\hat{p})}{(\hat{p}-\hat{p}_0)^m}
\end{equation}
and where $g(\hat{p})$ is analytic as $\hat{p}\rightarrow \hat{p}_0$. 
Nevertheless, we do expect very mild damping of the waves 
($\hat{\omega}_i \ll \hat{\omega}_r$) at least for 
the weakly relativistic case as discussed in \cite{LaingDiver2005}, where 
they find Im($\hat{\omega}$) to be the largest on the upper half of the curve 
that advances towards the cyclotron harmonics, and relatively 
much smaller on the lower half. It remains to be seen if 
damping is at all observed in the moderately relativistic case.  

The results presented in this article have important implications for
all astronomical objects where a magnetized hot pair plasma is
present, for example radio pulsars \cite{RudermanSutherland1975}, 
magnetars \cite{ThompsonDuncan1995}, pair-instability supernovae 
\cite{Fryeretal2001}.

\begin{acknowledgements}
   The authors thank the referee for useful comments. 
   J.S.H. would like to thank Declan Diver for useful conversations.
   The Natural Sciences and Engineering Research Council of Canada,
   Canadian Foundation for Innovation and the British Columbia
   Knowledge Development Fund supported this work. Correspondence and
   requests for materials should be addressed to
   J.S.H. (heyl@phas.ubc.ca). This research has made use of NASA’s
   Astrophysics Data System Bibliographic Services
\end{acknowledgements}

\end{document}